\documentclass[%
 reprint,
 amsmath,amssymb,
 aps,
 prb,
]{revtex4-1}
\usepackage{natbib}
\usepackage{epstopdf}
\usepackage{graphicx}
\usepackage{dcolumn}
\usepackage{bm}
\usepackage{xcolor}
\usepackage{ulem}
\usepackage{hyperref}
\usepackage{cleveref}

\begin{document}

\preprint{APS/123-QED}

\title{Modeling tunnel field effect transistors - from interface chemistry to non-idealities to circuit level performance}


\author{Sheikh Z. Ahmed }
\affiliation{
	Department of Electrical and Computer Engineering, University of Virginia, Charlottesville, Virginia 22904, USA
	}
\author{Yaohua Tan}
\affiliation{
	Department of Electrical and Computer Engineering, University of Virginia, Charlottesville, Virginia 22904, USA
	}
\author{Daniel S. Truesdell }
\affiliation{
	Department of Electrical and Computer Engineering, University of Virginia, Charlottesville, Virginia 22904, USA
	}
\author{Benton H. Calhoun }
\affiliation{
	Department of Electrical and Computer Engineering, University of Virginia, Charlottesville, Virginia 22904, USA
	}
\author{Avik W. Ghosh }
\affiliation{
	Department of Electrical and Computer Engineering, University of Virginia, Charlottesville, Virginia 22904, USA
	}

%


\date{\today}

\begin{abstract}
We present a quasi-analytical model for Tunnel Field Effect Transistors (TFETs) that includes the microscopic physics and chemistry of interfaces and non-idealities. 
The ballistic band-to-band tunneling current is calculated by modifying the well known Simmons equation for oxide tunneling, where we integrate the Wentzel-Kramers-Brillouin (WKB) tunneling current over the transverse modes. 
We extend the Simmons equation to finite temperature and non-rectangular barriers using a two-band model for the channel material and 
an analytical channel potential profile obtained from Poisson's equation.
The two-band model is parametrized first principles by calibrating with hybrid Density Functional Theory calculations, and extended to random alloys with a band unfolding technique.  Our quasi-analytical model shows quantitative agreement with ballistic quantum transport calculations. 
On top of the ballistic tunnel current we incorporate higher order processes arising at junctions coupling the bands, specifically interface trap assisted tunneling and Auger generation processes. Our results suggest that both processes significantly impact the off-state characteristics of the TFETs -  Auger in particular being present even for perfect interfaces. We show that our microscopic model can be used to quantify the  TFET performance on the atomistic interface quality. Finally, we use our simulations to quantify circuit level metrics such as energy consumption.
\end{abstract}

\pacs{Valid PACS appear here}
\maketitle


When it comes to the down-scaling of semiconductor transistors, Moore's Law has had a spectacular run, one that  has unfortunately reached its inevitable slow-down. 
For complementary metal oxide semiconductor (CMOS) devices, power dissipation has become a major bottleneck for digital applications~\cite{Bernstein10}, constrained ultimately by the fundamental Boltzmann limit of $k_BT\ln{10}/q \sim 60$ mV/decade that sets the steepness or subthreshold swing (SS) for the gate transfer characteristic \cite{Choi07} in conventional MOSFETs. To overcome the theoretical limit of SS for low power applications, novel transistor architectures such as tunneling FETs (TFET) \cite{TFETReview}, Mott Transition FETs \cite{shukla2015steep}, Graphene Klein Tunnel FETs \cite{sajjad2013manipulating} and Negative capacitance FETs \cite{NCFET}, have been proposed and widely investigated in the past decade. Among those novel device architectures, the most widely studied are TFETs that operate on the abrupt opening of gate controlled transmission channels through band-to-band (Zener) tunneling across a reverse bias pn junction \cite{Ghosh15,Seabaugh14}. Unfortunately, none of the reported TFETs in the literature, to the best of our knowledge, approach the dual needs of high current for fast speed and a low subthreshold swing over several decades from ON to OFF in order to get low voltage operation and thereby low dynamic power dissipation.

A number of TFET designs across a wide variety of device structures and materials have been  investigated theoretically \cite{Seabaugh14}. The usual approach to studying the detailed physics of these devices is atomistic quantum transport based on Non-equilibrium Green's Functions (NEGF) \cite{luisier2009performance,Avci12,Huang,Long}. Atomistic quantum transport modelings are capable of revealing physical insights but are challenging in many aspects. For instance, many quantum transport studies use empirical models such as $k\cdot p$ and tight-binding models that are numerically-efficient but are fitted to bulk properties and have questionable transferability at surfaces and interfaces. 
Additionally, the intrinsic numerical complexity of atomistic quantum transport modeling causes them to become excessively time-consuming for systems with realistic device size. As a result, most studies using quantum transport modeling are limited to ballistic transport of small-sized devices which predicted much lower off-current and SS compared to experimental results~\cite{Seabaugh14}.
Quantum transport calculations with scatterings are even more computationally expensive \cite{luisier2009atomistic}, and are thus more size constrained.
This high computational burden makes it difficult to consider processes such as trap and defect-assisted tunneling \cite{SajjadTraps}, electron-phonon scattering,\cite{carrillo2016effect} and electron-electron interactions such as Auger generation \cite{Teherani_E3S}. Such processes have zeroth order effects on device performance, for instance by introducing leakage currents that raise the current floor and limit the SS of heterostructure TFET devices. For a fast  intuitive way to estimate the TFET device physics including crucial high-order processes, a proper physics based compact model is needed. Accurate quasi-analytical modeling is also needed to bridge numerical modeling and experimental data with circuit-level studies and simulations. 

The central component of any TFET structure is band to band tunneling (BTBT) across a new channel that opens at the source end of a p-i-n junction.
To model BTBT current, many existing analytical TFET models\cite{Zhang14,Vishnoi14,Vishnoi15,Lu15} use Kane's approach \cite{KANE}, which uses WKB approximation to estimate the tunneling current through triangular barriers within a simple effective mass approximation. {However, these models typically do not explicitly include the Fermi tails but incorporate them through fitting functions.}
Moreover, Kane's model requires extra model parameters to compensate for errors cause by the over-simplified bands and 1D electrostatics. There is thus a pressing need for a chemistry-based analytical model based on a proper tunneling equation that accounts for multiple transverse modes, avoids `fudge factors' and is rooted in chemical modeling and realistic electrostatics, used thereafter to calculate temperature-dependent BTBT current across complicated junctions~\cite{Simmons}. To understand the discrepancies between theoretical ballistic current and experimental current, higher order effects such as defects have been included (albeit sparingly) in previous TFET analytical models. These include Trap Assisted Tunneling (TAT) studies by Sajjad {\it{et al.}}~\cite{SajjadTraps}, and Auger generation in perpendicular TFETs by Teherani {\it{et al.}}~\cite{Teherani}. 

In this work, we present a physics-based analytical model for planar TFETs. 
This analytical model makes use of the potential obtained by solving the pseudo-2D Poisson's equation. 
A simplified two-band model is used to describe the electronic properties of the channel, source and drain materials. The material parameters are extracted from tight binding band structures that have been calibrated with first principles band structures and wave functions. Using the two-band model and the approximated potential, ballistic band to band tunneling is calculated using modified Simmons equation. On top of the ballistic model, we introduce the impact of trap assisted tunneling and Auger effect, and quantify their impact at the circuit level. 

\section{A quasi-analytical Model}
\begin{figure}[h] 
\centering
\includegraphics[width=0.45\textwidth]{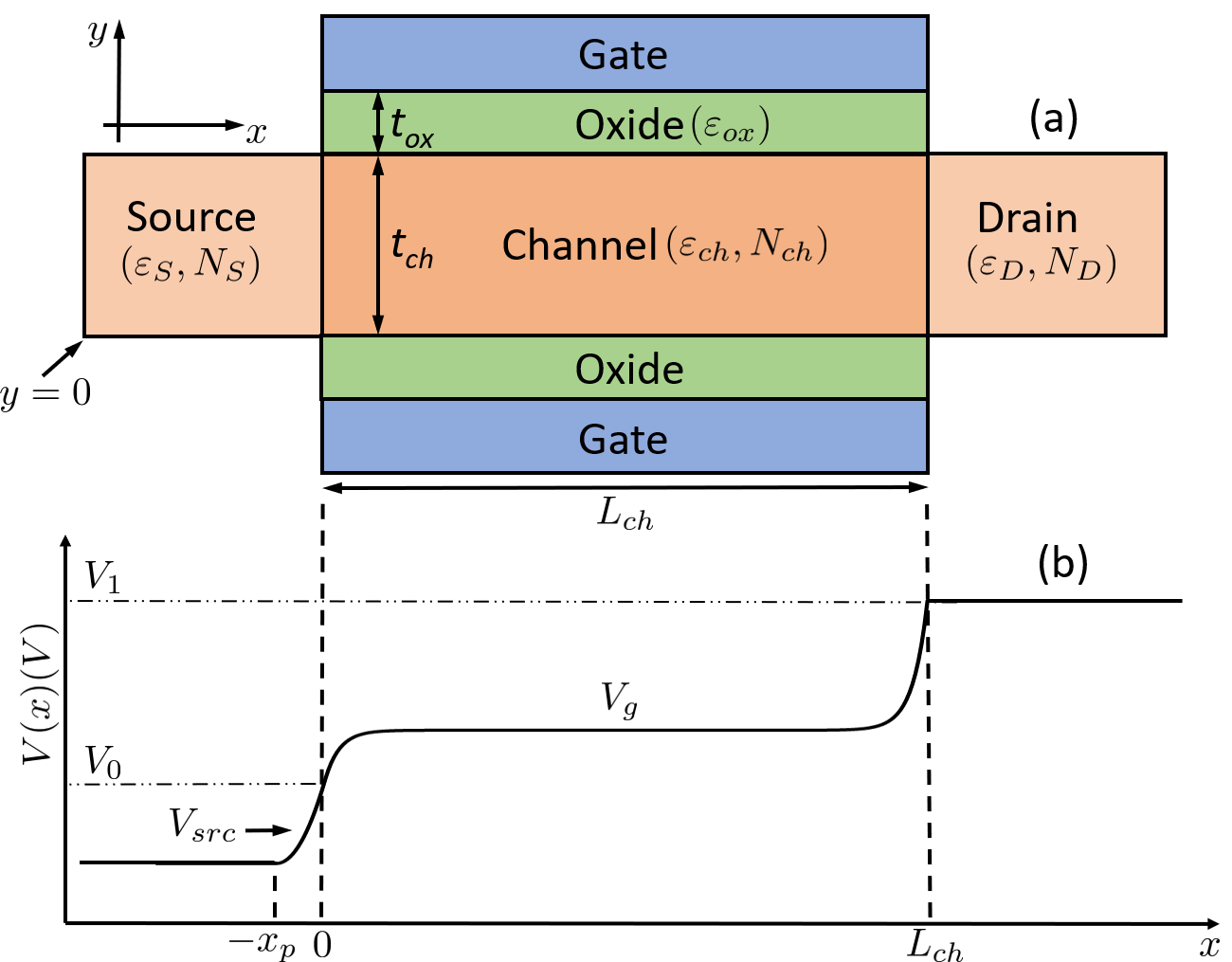}
\caption{ Planar TFET structure considered in this work(a) and potential in the device(b). In the analytical model, 2D Poisson's equation in the channel region is solved and approximated by analytical equations.  
}\label{fig:possion_regions}
\end{figure}

The geometry of a $n$-type double-gated TFET is shown in Fig.~\ref{fig:possion_regions} (a). The source, channel and drain regions are $p^+$, $i$ and $n^+$ doped, respectively. The doping concentrations are $N_S$ for source, $N_{ch}$ for channel, and $N_D$ for drain. The channel region is rectangular with a width of $t_{Ch}$ and channel length of $L_{ch}$. The gate oxide has a thickness of $t_{ox}$. 
The dielectric constants for source, drain, channel and gate oxides are $\varepsilon_{S}$, $\varepsilon_{D}$,  $\varepsilon_{ch}$ and  $\varepsilon_{ox}$. In this work, we consider both homojunction and heterojunction TFETs - the former targeting a pristine interface for low OFF current while the latter allowing a thin tunnel barrier across a staggered gap (Type II) junction for large ON current. The homojunction TFET has an In$_{0.53}$Ga$_{0.47}$As channel. The heterojunction TFET has a GaSb source and an InAs channel/drain. 

\subsection{TFET Surface Potential}\label{sec:potential}

For our device structure the surface potential, shown in Fig.~\ref{fig:possion_regions} (b), is obtained by solving Poisson's equation with appropriate boundary conditions. 
Since the tunneling current in a TFET is dominated by the source/channel junction, the potential at that junction is critical.
{While we know the dopings in the different regions, the potential in the channel and thus the voltage barrier across the source-channel region is complicated, and needs to be computed including the gate and drain capacitances. We will now simplify the solution to 2D Poisson's equation with suitable boundary conditions to extract the 1D channel potential at the gate-channel interfaces, culminating in Eq.~\ref{evch}}.

The surface potential $V_0$ at the boundary of the the heavily doped source region can be calculated by solving 1D Poisson's equation with a uniform doping concentration $N_S$ and dielectric constant $\epsilon_S$, assuming a homogeneous potential along the vertical $y$ direction. 
The solution to this equation can be written as $ V_{src}(x)={qN_S}{}(x+x_p)^2/2\epsilon_{S}$ where $x_p$ is the depletion width at the junction along the x-direction as shown in Fig. \ref{fig:possion_regions} (b). This gives us one boundary condition at the source/channel interface, the other at the drain/channel interface {is obtained from the offset in the local quasi-Fermi levels between source and drain}
\begin{eqnarray}
    V_0 &=&V{src}(0) = \frac{qN_S}{2\epsilon_{S}}x_p^2\\
    V_1 &=&\frac{kT}{q}\ln\left(\frac{N_SN_D}{n_{iS}n_{iD}}\right) +V_{DS} + \Delta E,
    \label{ebcs}
\end{eqnarray}
where $N_S,N_D$ represent the doping concentrations in the source and drain regions, and $n_{iS}$ and $n_{iD}$ are the intrinsic carrier concentrations of the source and drain materials, while $\Delta E$ is the band offset between source and drain materials. $V_0$ is the solution of 1D Poisson's equation at the source and will need to be estimated shortly. For $V_1$, we assume the  potential is  constant in the drain region. Compared with the rigorous solution in the drain, this approximation leads to negligible difference in the current because the band-to-band tunneling occurs at the source-channel interface. 

For the potential in the channel, a pseudo-2D Poisson's equation is solved in the rectangular channel region shown in Fig.~\ref{fig:possion_regions} (a). It is assumed that mobile charge carriers do not effect the electrostatics of the device \cite{Shen} compared to the fixed charges (dopants). The 2D Poisson's equation in the channel region can be written as 
\begin{equation}\label{eq:2D_poisson_channel}
    \frac{\partial^2V(x,y)}{\partial x^2}+\frac{\partial^2V(x,y)}{\partial y^2}=\frac{qN_{ch}}{\epsilon_{ch}}
\end{equation}
where $V(x,y)$ is the electrostatic potential of the region, $N_{ch}$ is the effective doping and $\epsilon_{ch}$ is the dielectric constant of the material. Considering a parabolic variation of the potential in the y-direction ($y=0$ being the bottom channel-gate interface and $y=t_{ch}$ the top) the 2D potential can be approximated by the second order polynomial in $y$ \cite{young}, 
\begin{equation}\label{e4}
V(x,y)=a_0(x)+a_1(x)y+a_2(x)y^2
\end{equation}
We first {use the continuity of potential and displacement field in the y direction to convert the 2D Poisson equation into an equivalent 1D equation for the channel potential $V_{ch}(x)$. The boundary conditions are set by the gate potential $V_g$ and field at the lower $y=0$ and upper $y=t_{ch}$ gate-channel interfaces}
\begin{eqnarray}
 V \left( x, 0\right)&=& V_{ch} \left(x\right) \nonumber\\
 V \left( x, t_{ch}\right)&=& V_{ch} \left(x\right)  \nonumber \\
 E_y \left( x, 0\right)  &=& -\frac{\eta}{t_{ch}}\left(V_G - V_{ch}(x) \right)   \nonumber  \\
 E_y \left( x, t_{ch}\right)&=&-\frac{\eta}{t_{ch}}\left(V_{ch}(x)-V_G \right) \label{evbcs}
\end{eqnarray}
The parameter $\eta=C_{ox}/C_{ch}$ represents the ratio between the gate capacitance $C_{ox}$ and the channel capacitance $C_{ch}=\epsilon_{ch}/t_{ch}$. The gate potential $V_g$ is referenced with respect to the flatband condition $V_G=V_{GS}-V_{fb}$, where $V_{fb} = \phi_m+\chi+E_g/2$,  $\phi_m$, $\chi$ and $E_g$ representing the gate metal work function, electron affinity and the bandgap of the channel material, respectively.  

By applying these four vertical boundary conditions at the gate/channel interfaces, {we can find the coefficients in} equation (\ref{e4}). {Since the electric fields are largest at the channel-gate interfaces $y=0, ~t_{ch}$, the tunneling electrons are preferably attracted to those interfaces. We thereafter  focus on the channel potential   $V_{ch}(x) = V(x,y=0)$. Substituting in Eq.~\ref{eq:2D_poisson_channel}, we find that $V_{ch}(x)$ satisfies a 1D Poisson equation} 
\begin{equation}\label{eq:1d_poisson}
V_{ch}^{''}\left(x\right) - k^2  V_{ch}\left(x\right) = - k^2V_g
\end{equation}
with
\begin{eqnarray}
 k &=&\sqrt{2\eta/t_{ch}^2}   \nonumber \\
 k^2V_g &=& k^2V_G -\frac{qN_{ch}}{\epsilon_{ch}}
 \label{evd}
\end{eqnarray}
Here, the characteristic length in the channel region is given by 1/$k$ and $V_g$ represents the solution of 1D approximation of Poisson's equation using the long-channel approximation.  
The solution $V_{ch}$ 
can be written as\cite{Bardon} 
\begin{equation}
V_{ch}(x) =be^{kx}+ce^{-kx}+V_{g} 
\label{evch}
\end{equation}
{with boundary conditions $V_{0,1}$ at the two ends (Eq.~\ref{ebcs}), with $V_0$ still unknown.}
\begin{eqnarray}
 b&=&\frac{1}{2\sinh{(kL_{ch})}}\left(-V_0e^{-kL_{ch}}-V_g(1-e^{-kL_{ch}})+V_1\right)\nonumber\\
 c&=&\frac{1}{2\sinh{(kL_{ch})}}\left(V_0e^{kL_{ch}}+V_gd(1-e^{kL_{ch}})-V_1\right)
 \label{ebc}
\end{eqnarray}
The {related} unknown variable is $x_p$, the width of the depletion region width in the source.
$x_p$ can be obtained using the continuity of the displacement field at source/channel interface 
\begin{equation}\label{eq:boundary_matching}
    \epsilon_{S}\frac{d V_{src}}{dx}=\epsilon_{ch}\frac{dV_{ch}}{dx}.
\end{equation}
By substituting $V_{ch}(x)$ from Eq.~\ref{evch} and the form of $V_{src}(x)$ discussed before Eq.~\ref{ebcs},  {we get a nonlinear equation in $V_0$ or equivalently a quadratic equation in $x_p$ with a positive solution} 
\begin{equation}
    x_p=\frac{-1+\sqrt{1+{2{P}}/{qN_S\epsilon_S}}}{Q}
    \label{ebcs2}
\end{equation}
where,
\begin{gather}
    P=\epsilon_{ch}^2k^2\coth{(kL_{ch})}\left[V_g\coth{(kL_{ch})}
    -\frac{V_g-V_1}{\sinh{(kL_{ch})}}\right]\nonumber
\end{gather}
\begin{equation}
    Q={\frac{\epsilon_{ch}k\coth{(kL_{ch})}}{\epsilon_S}}
    \label{epq}
\end{equation}
{Our key equation is thus the channel potential
(Eq.~\ref{evch}), with coefficients $b$ and $c$ from Eq.~\ref{ebc}. The long channel potential $V_g$ is related to the applied gate bias and the doping through Eq.~\ref{evd} and the definition of $V_G$ after Eq.~\ref{evbcs}. The channel potentials at the two ends $V_0$ and $V_1$ are obtained from Eqs.~\ref{ebcs}, \ref{ebcs2} and \ref{epq}}. 

The potential model presented in this section can be applied to both homojunction TFET and heterojunction TFETs. {For a long channel, $kL_{ch} \gg 1$, we get $x_p \approx \sqrt{2\epsilon_SV_g/qN_S}$, as we expect for depletion widths across conventional PN junctions with $V_g$ replacing the built-in potential. The channel potential $V_{ch}(x) \approx V_0 + (V_1-V_0)\exp{k(x-L_{ch})}$ with $V_0 \approx V_g$, meaning that the potential stays pinned to $V_g$ for much of the channel length and switches to $V_1$ only within a distance $\sim 1/k$ of the drain end. For short channels $kL_{ch} \ll 1$, $x_P \approx \sqrt{2\epsilon_SV_1/qN_S}$ and $V_{ch}(x) \approx V_0 + (V_1-V_0)x/L_{ch}$.}

\begin{figure}[h]
\includegraphics[width=0.45\textwidth]{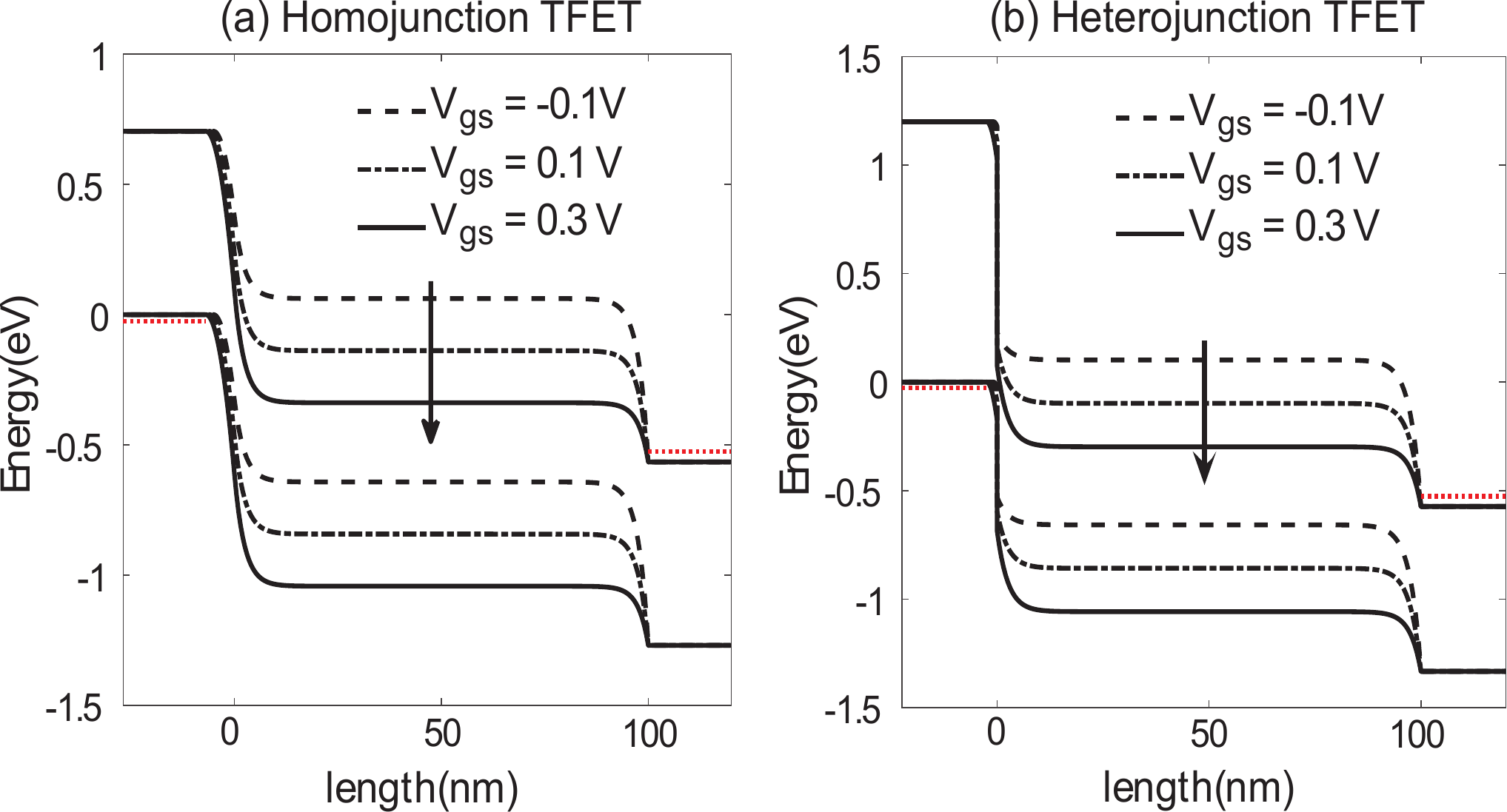}
\caption{ Band diagram of the ON and OFF states in (a) homojunction  and (b) heterojunction TFETs. The source region has $x<0~\textrm{nm}$, drain region has $x>100~ \textrm{nm}$, and the channel region  has $0~ \textrm{nm}\leq x \leq 100 ~\textrm{nm}$.  
}\label{fig:potential_homo_hetero}
\end{figure}

Fig.~\ref{fig:potential_homo_hetero} shows the band diagrams of a homojunction and a heterojunction TFET under different gate bias conditions. We assume both TFETs have 100~nm channel length. Material parameters are extracted from previous Non-Equilibrium Green's function calculations. The homojunction TFET has InGaAs as channel with a band gap of 0.74~eV ~\cite{Avci12}, and the heterojunction TFET has GaSb as source with a 1.2~eV band gap ~\cite{Huang} and InAs as channel with a band gap of 0.76~eV ~\cite{Long}.

\subsection{Two-band Model for the junction}\label{sec:band_model}
\begin{figure}[h]
\includegraphics[width=0.45\textwidth]{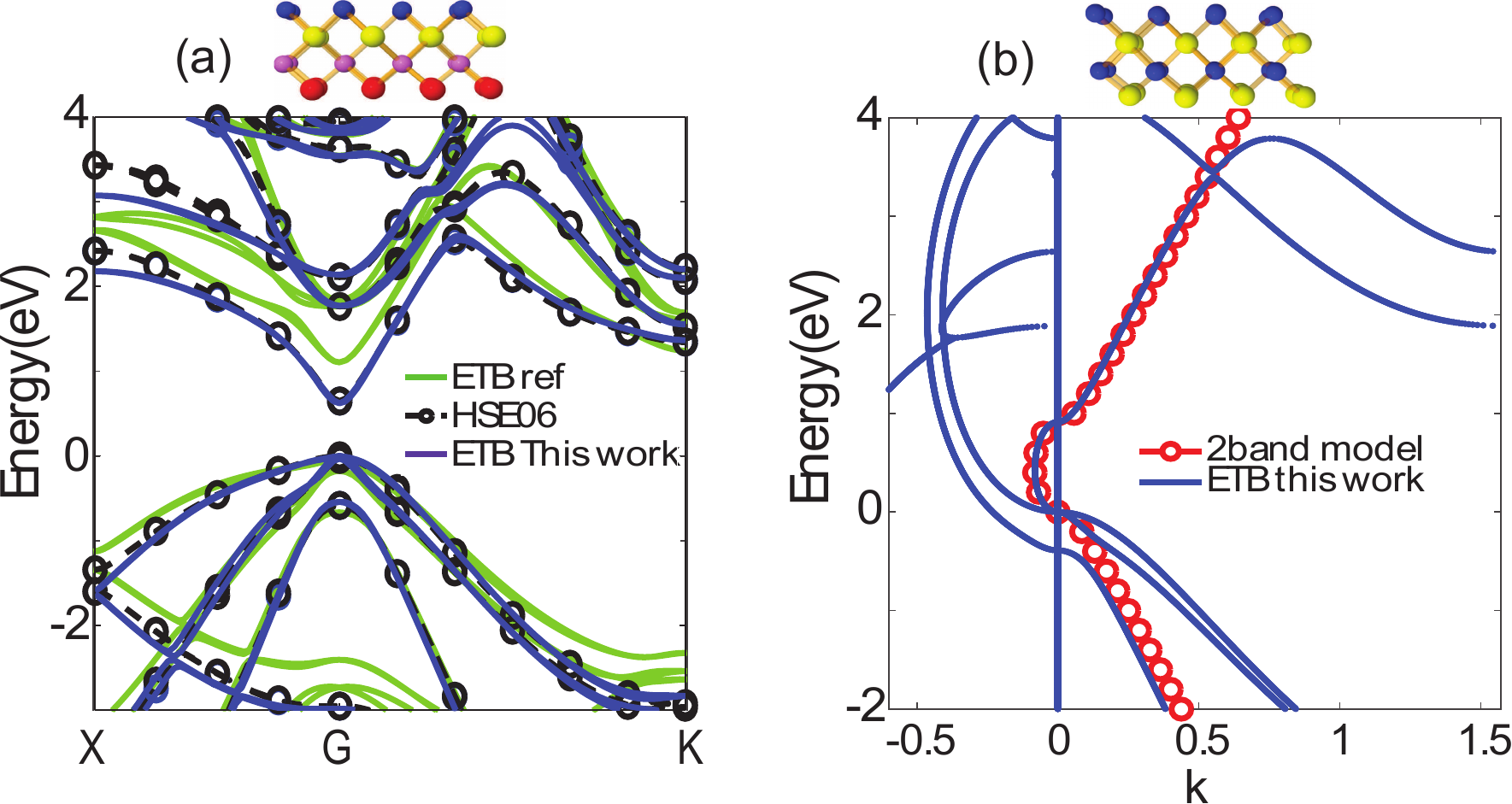}
\caption{ (a) Comparison of band structure of InAsGaSb superlattice. The tight binding calculation (blue lines) in this work agree well with the first principle band structure with hybrid functional calculation (HSE06) in black dotted lines. While the tight binding calculation using previous parameters(green line) show large discrepancies compared with first principle calculations. (b) Real and complex band structure of strained InAs two-band model vs tight binding. By adjusting the parameters of the two-band model, both the real bands and complex bands from the two-band model agree well with the tight binding model.}\label{fig:tightbinding_vs_DFT_2bandmodel}
\end{figure}

To model band to band tunneling correctly, a single band effective mass model is clearly insufficient. We use a simplified two-band $k\cdot p$ model which can generate more accurate real and complex bands for the direct band gap III-V group materials considered in this work. 
In this two-band model, the bands considered represent the conduction band and light hole band at the $\Gamma$ point, critical for the band to band tunneling process. 
For III-V materials, the complex band connecting conduction band ($\Gamma$) and light hole band has the smallest imaginary wave vector, as shown in Fig. \ref{fig:tightbinding_vs_DFT_2bandmodel} (b). Carriers tunneling from the light hole band to the lowest conduction band across the junction clearly dominate the  current. The heavy hole and split-off bands connected to higher conduction bands have complex bands with much larger imaginary wave vectors. These bands decay much faster in real space and can thus be ignored. The two-band model in this work can be written as
\begin{equation}
H(\mathbf{k}) = \begin{bmatrix}
E_c(\mathbf{k}_{\parallel}) & A k_x\\ 
A k_x & E_v(\mathbf{k}_{\parallel})
\end{bmatrix},
\end{equation}
The equation generates two parabolic bands with a common {tunneling} effective mass $m^*$ that is obtained by setting 
${A^2}/{(E_c-E_v)}={\hbar^2}/{2m^*}$. {While this model ignores the separate masses for conduction and valence band within a single material, we assume band-to-band tunneling occurs from a single light hole valence band in the source to a single conduction band in the channel, which means a single mass separately set in each material suffices to capture the dominant TFET current}.  
The $E_c$ and $E_v$ dispersions are set by the potential we just worked out, shown in Fig. \ref{fig:possion_regions}b.
\begin{eqnarray}
E_c(\mathbf{k}_{\parallel},x) &=& E_c-qV(x)+{\hbar^2}k_{\parallel}^2/{2m^*}\nonumber\\ E_v(\mathbf{k}_{\parallel},x) &=& E_v-qV(x)-{\hbar^2}k_{\parallel}^2/{2m^*}\nonumber\\
V(x) &=& \begin{cases}
V_{src}(x), & x \leq x_0  \\
V_{ch}(x), &  x_0 < x < L_{ch} \\
V_1, & x > x_0
\end{cases}
\end{eqnarray} 
The required material parameters for the two-band model are the electron {tunneling} effective mass $m^*$ and band edges $E_c$ and $E_v$. The bold $\mathbf{k}$'s correspond to vectors, and italic $k$'s correspond to scalars.

With the two-band model, for a given energy $E$, the $k_x$ can be calculated analytically as
\begin{equation}\label{eq:k_x}
k_x \left(\mathbf{k}_{\parallel},x\right) =\pm\frac{\sqrt{(E-\bar{E})^2-\Delta^2}}{A}
\end{equation}
with $\bar{E} = \left(E_c(\mathbf{k}_{\parallel},x) + E_v(\mathbf{k}_{\parallel},x)\right)/2$ and $\Delta = \left(E_c(\mathbf{k}_{\parallel},x) - E_v(\mathbf{k}_{\parallel},x)\right)/2$. 
This expression works for any given energy $E$, and can generate both real and complex k. 
{For a given E, all possible $k_x$s are calculated within the first brillouin zone of $k_\parallel$ and their transmissions will eventually be summed up to get the total transmission over perpendicular states} 

\subsection{Accurate parametrization and band-unfolding at the junction}\label{sec:tight binding}
In this work, we extract the material parameters from tight binding calculations.
It should be emphasized that the accuracy of tight binding has a significant impact on the results, since the tunneling current depends exponentially on the tunneling effective mass, which is ultimately a hybrid between the bulk light hole and conduction band effective masses on its sides.
We employ a tight binding model that has been carefully calibrated not only with band structure but also wavefunctions based on experiments as well as high accuracy first principles calculations \cite{tan2015tight,tan2016transferable}.
 We fit our tight binding parameters with Density Functional Theory (DFT) within the HSE06 hybrid functional\cite{heyd2003hybrid} approximation that is known to generate accurate bandstructure of semiconductors matching experiments. In the past, we demonstrated that one way to make tight-binding transferrable was to employ non-orthogonal basis sets to calibrate bond overlaps in Extended H\"uckel Theory \cite{eht1,eht2}. Here we employ an alternate way to endow orthogonal tight binding with transferrability between bulk geometries and systems with strain and interfaces, by additionally matching the radial wavefunctions with DFT. 

Fig.~\ref{fig:tightbinding_vs_DFT_2bandmodel} (a) shows excellent agreement between our tight binding calculations and hybrid functional(HSE06) results for systems with interfaces, in this case an ultra small $\textrm{InAs/GaSb}$ superlattice. In comparison,  tight binding calculations using previous  parameters \cite{jancu1998empirical} show an obvious discrepancy, because they are extracted by fitting to bulk InAs and GaSb bands without considering physical insights from wave functions, and without calibration to interfaces. The comparison suggests that the tight binding we developed has much better transferrability for III-V superlattices and alloys. 
Fig.~\ref{fig:tightbinding_vs_DFT_2bandmodel} (b) shows how well our simplified two-band model matches the conduction band, light hole band and the complex bands from tight binding calculations. {For comparison  studies with past NEGF calculations, we retain parameters $m^*$ and $E_g$ from previous work to keep the benchmarking standards the same}.

\begin{figure}[h]
\includegraphics[width=0.45\textwidth]{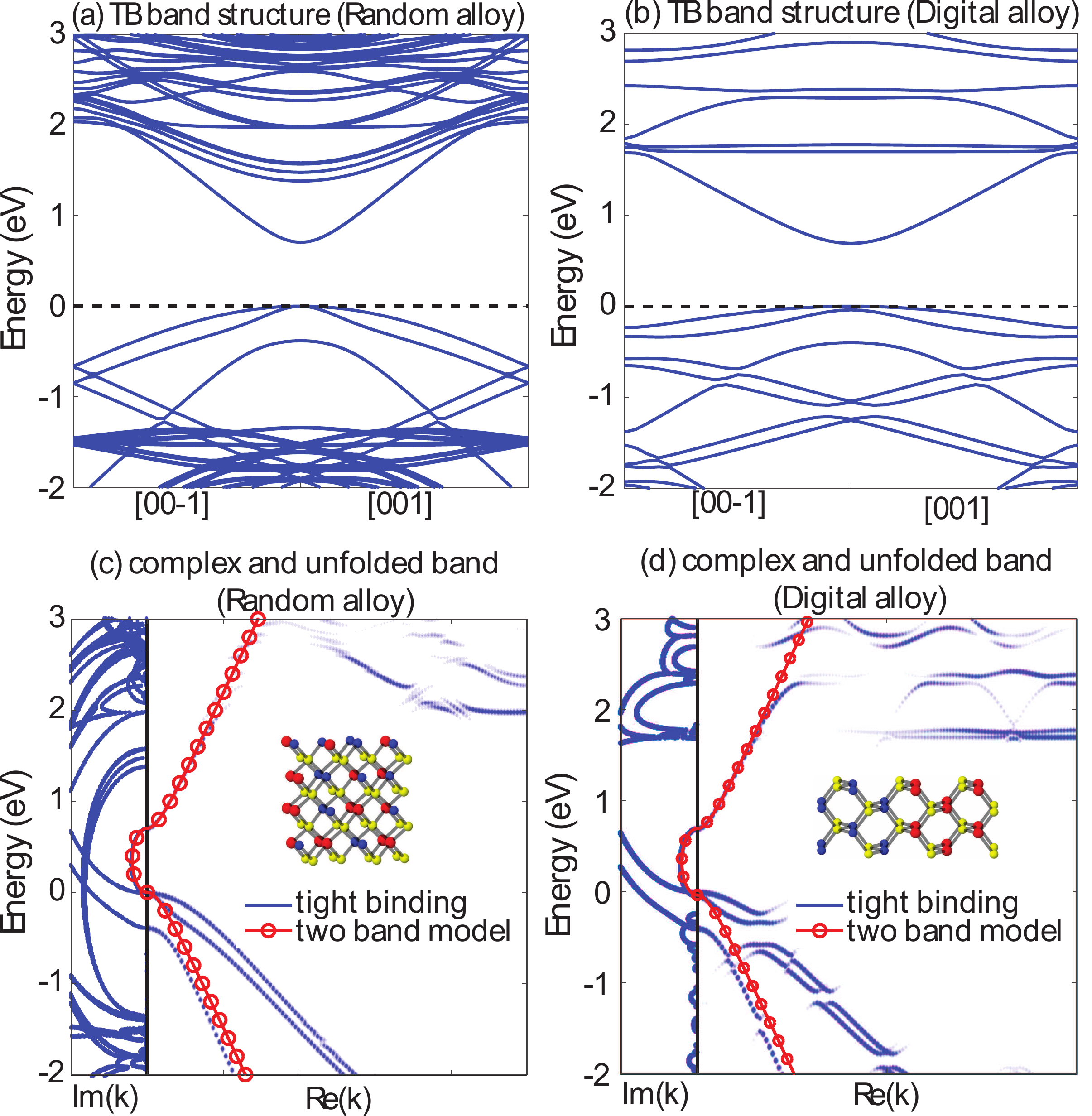}
\caption{ Real and complex band structure of strained $\textrm{In}_{0.5}\textrm{Ga}_{0.5}\textrm{As}$ alloys with two-band model and tight binding model. (a) original tight binding band structure and (c) unfolded real and complex band structure of a Random alloy. (b) original tight binding band structure and (d) unfolded real and complex band structure of a Digital alloy. By adjusting the parameters of the two-band model, both the real bands and complex bands from the two-band model agree well with the tight binding model. }\label{fig:model_complex_bands_random_digital_alloy}
\end{figure}

For the alloy $\textrm{In}_{0.5}\textrm{Ga}_{0.5}\textrm{As}$ in the Homo-junction TFET, we studied two different cases - namely a random alloy and a digital alloy. The tight binding band structures and two-band model band structures for the two geometries are shown in Fig.~\ref{fig:model_complex_bands_random_digital_alloy}. It can be seen from (a) and (b) that the tight binding band structures for a random alloy and a digital alloy share little resemblance with each other, because of their vastly different unit cell sizes. {Here, we only consider a specific instance of a random alloy whereas for a practical device, the TB band structure must be obtained by doing a Monte Carlo averaging to account for the distribution of the defects.} In order to make a meaningful comparison between random and digital alloy in (c) and (d), we used the technique of band unfolding \cite{tan2016first,Boykin_BZ_unfolding_SiGe,Boykin_BZ_unfolding_alloy} to simplify the real bands. {In this technique, the wavefunctions of the supercell in the alloy Brillouin zone are fourier decomposed and the Fourier coefficients used to generate the wavefunction probabilities at the corresponding $k$ and energy points.}
We see how some of the high energy conduction bands and low energy valence bands are unfolded back to the Brillouin zone of the primitive zincblende unit cell. It can be seen that the unfolded direct conduction band and valence bands in digital and random alloy have similar band profiles. Notably, the digital alloy creates broken bands with minigaps due to coherent destructive interference, suggesting a strong interaction among the bands. 
To visualize the tunneling properties of  InGaAs,the complex bands in a random and digital alloy are also shown in (c) and (d). More complex branches can be seen in the random alloy than the digital alloy, corresponding to the fact that there are more bands for the former. However, the two band model shows a much simpler band profile with only  the most important complex band captured. By adjusting the parameters of the two-band model, we can match the direct conduction band, dominant light hole band and the complex branch connecting the two band-edges for both digital and random alloys. The agreement implies the two band model is a good approximation to model band to band tunneling even for III-V alloy materials.  

\subsection{BTBT Current Model}
\begin{figure}[h] 
\centering
\includegraphics[width=0.45\textwidth]{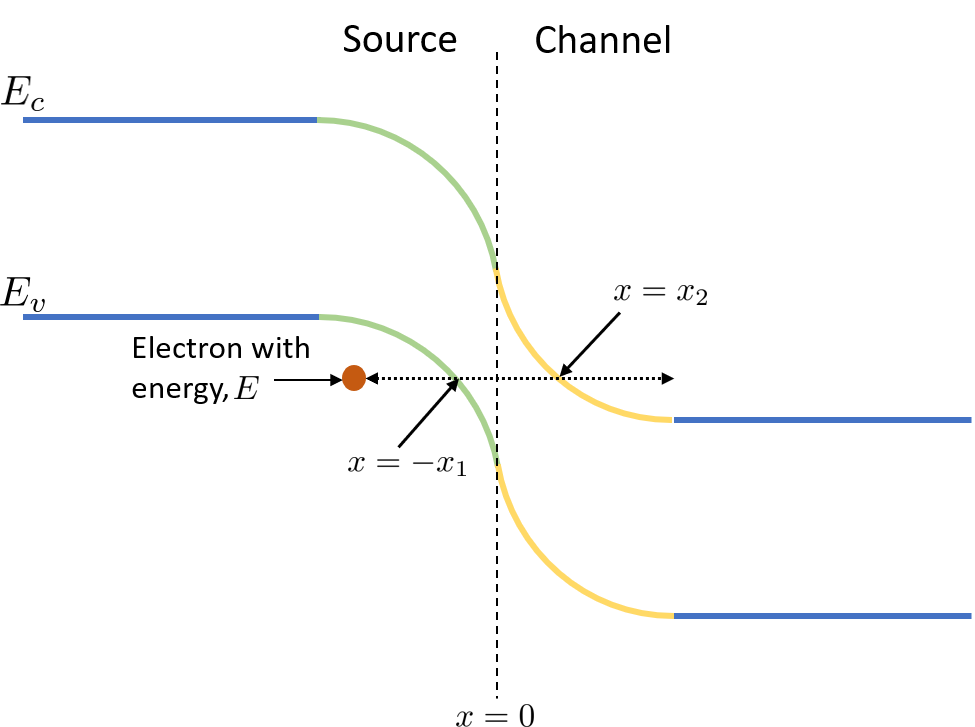}
\caption{ Band diagram of the tunneling junction. The distance between $-x_1$ and $x_2$ is the tunneling width for  the electron shown.  }\label{fig:tunneling_regions}
\end{figure}
With the k vectors set by the two-band model, the BTBT current is ready to be calculated. The Simmons model has been very successful in the chemistry literature in quantatively describing the tunneling current through a thin film. This model approximates the WKB tunneling current by Taylor expanding the barrier profile in the exponent around a rectangular shape and then summing over a continuum of transverse modes
 \cite{Simmons}. 
We modified the Simmons equation to account for a non-rectangular shape in the current integration over perpendicular $k$,  and also retained a finite temperature dependence that sets the subthreshold swing for switching.  The modified Simmons equation {arises from Landauer equation} 
\begin{equation}
    I=\frac{q}{h}\int_{E_{FD}}^{E_{v,src}} T\left(E\right)\left[ f_S(E)-f_D(E)\right]dE
\end{equation}
Here the $k_{\parallel}$ is calculated by equation (\ref{eq:k_x}) in section \ref{sec:band_model}, and the band to band tunneling probability $T\left(k_{\parallel},E\right)$ is estimated by WKB approximation 
\begin{equation}
    T\left(k_{||},E\right) = \left( 1-R \right) \left|\exp{\left(-2i \int k_x\left(x\right) dx \right) }\right| 
\end{equation}
where,
\begin{equation}
   \int k_x\left(x\right) dx= \int_{-x_1}^{0} k_x\left(x\right) dx+ \int_{0}^{x_2} k_x\left(x\right) dx \nonumber
\end{equation}
In this equation, the $T\left(k_{||},E\right)=1$ for any real $k_x$, and $T\left(k_{||},E\right)<1$ for complex $\kappa_x$ corresponding to the process of tunneling through a barrier. The boundary conditions for determing $x_1$ and $x_2$ are $E-E_v(-x_1)=0$ and $E-E_c(x_2)=0$, respectively. The band diagram of the tunneling region is shown in Fig.~\ref{fig:tunneling_regions}. To take account of added reflections at the interface  not accounted for in the WKB approximation (arising from the kinetic energy pre-factor in the semiclassical approximation), we introduce a phenomenological correction factor $1-R<1$ with $R$ the reflection coefficient at the interface. The reflection $R$ is manually adjusted to mimic the effect of interface reflection and achieve better agreement with the reference I-V. The integral is separated into the source and channel regions in order to obtain an analytic expression.

To obtain a closed-form solution of the source region integral, the source potential is at first approximated using linearization applied at the point $x=-x_1/2$ 

\begin{equation}
V_{src,li}\left(x\right) = \frac{qN_S}{\epsilon_S} \left[ \frac{1}{2} \left(x_p-\frac{x_1}{2} \right)^2+ \left(x_p-\frac{x_1}{2} \right) \left(x+\frac{x_1}{2}\right) \right] 
\end{equation} 

\begin{equation} \label{eq:wkb_source}          
      V_{src,li}\left(x\right) = V_{src}\left(-\frac{x_1}{2}\right)+V_{src}^{'} \left(-\frac{x_1}{2}\right)\left(x+\frac{x_1}{2}\right)
\end{equation}
Using equation (\ref{eq:wkb_source}) the result of the indefinite source integral can be written as 
\begin{eqnarray}
    \int k_x\left(x\right) dx=\frac{-\epsilon_S}{2AqN_S \left(x_p-\frac{x_1}{2}\right)}\left[\tilde{E}\sqrt{\tilde{E}^2-\Delta^2} \right. \nonumber\\
 \left.  -\Delta^2 \log \left( \tilde{E} + \sqrt{\tilde{E}^2-\Delta^2}\right) \right] \nonumber
\end{eqnarray}
where,
\begin{equation}
    \tilde{E}=E-\frac{E_c+E_v}{2}+V_{src,li}\left(x \right)\nonumber
\end{equation}
The channel region integral is approximated by assuming that the $be^{kx}$ term in the channel potential is negligible near the source/channel junction 
{where the potential profile is starting to saturate. Thus, the solution of the indefinite integral can be written as
\begin{eqnarray}
     \int k_x\left(x\right) dx= -\frac{1}{kA} \left[-\sqrt{\tilde{E}^2-\Delta^2}+\sqrt{C_1^2-\Delta^2}kx \right. \nonumber\\
     -C_1 \log\left(\tilde{E}+\sqrt{\tilde{E}^2-\Delta^2}\right)\nonumber\\
     +\sqrt{C_1^2-\Delta^2} \log \left(C_1^2-\Delta^2+C_1ce^{-kx} \right. \nonumber\\
  \left. \left.  +\sqrt{C_1^2-\Delta^2}\sqrt{\tilde{E}^2-\Delta^2}\right) \right] \nonumber\\
\end{eqnarray}
where,
\begin{equation}
    C_1=E-\frac{E_c+E_v}{2}+V_d
\end{equation}
The total transmission $T(E)$ is calculated by integrating the tunneling probability over all $k$-states {parallel to the interface} in the first Brillouin zone {over a 2D circular phase space}
\begin{equation}\label{eq:perp_k_integral}
    T(E)=2\pi \int_0^{k_{||max}}k_{||} T\left(k_{||},E\right) dk_{||}
\end{equation}
A fitting equation is used to calculate the value of $k_{||max}$ since for large values of $k_{||max}$ the two-band model deviates from the actual bandstructure. Here, $k_{||max}=\left|c_1 E+c_2\right|$. The constants $c_1=5.6\times10^8$ $m^{-1} eV^{-1}$, $ c_2=5.1\times10^8$ $m^{-1}$ and energy $E$ has units of $eV$. The integral in equation (\ref{eq:perp_k_integral}) does not have an analytical solution. Thus, an approximate solution for $T(E)$ must be obtained by

\begin{equation}\label{eq:perp_k_appr}
    T(E)=2\pi \sum_{n=0}^3 \int_{k_{||n}}^{k_{||n+1}} f\left(k_{||}\right) dk_{||} 
\end{equation}
where $k_{||n}=nk_{||max}/{4}$ and $ f\left(k_{||}\right)=k_{||}T\left(k_{||},E\right)$. The integral in equation (\ref{eq:perp_k_appr}) can be computed using Simpson's $3/8$ rule.
\begin{eqnarray}
  \int_{k_{||n}}^{k_{||n+1}} f\left(k_{||}\right) dk_{||}=\frac{k_{||n+1}-k_{||n}}{8}
    \left[3f\left(\frac{2k_{||n}+k_{||n+1}}{3}\right) \right. \nonumber\\
    \left.+3f\left(\frac{k_{||n}+2k_{||n+1}}{3}\right)+f\left(k_{||n}\right)+f\left(k_{||n+1}\right)\right] \nonumber\\
    \end{eqnarray}
\subsection{Trap Assisted Tunneling}
To account for critical high order effects near the source-channel junction, a trap assisted tunneling (TAT) process is included in our model. Due to the existence of defects near a material interface, intermediate energy levels known as trap states form a quasi-continuous density of states in the band gap.
Electrons can jump from the source valence band into the channel conduction band though these trap states by exchanging energy with optical phonons. This undesired flow of electrons creates a leakage current which have many adverse effects on TFET performance, namely higher off-current and higher SS. The trap current per unit width, $I_{TAT}$, is calculated using {by using a Fowler-Nordheim type tunneling through a tilted barrier around the trap. The Shockley-Reed-Hall generation rate is given by\cite{SajjadTraps}  }
\begin{equation}
    G=\int \frac{\sigma_n \sigma_p v_{th}\left(n_i^2-np\right)}{\sigma_n \frac{n+n_1}{1+\Gamma_p}+\sigma_p\frac{p+p_1}{1+\Gamma_n}} D_{it} dE
\end{equation}
where $\Gamma$ describes the electric field-enhancement of the trap assisted tunneling and thermionic emission processes, $n_i$ is the intrinsic charge concentration, $n, p$ are the electron and hole densities in the conduction and valence band, $D_{it}$ is the interface trap density, assumed primarily mid-gap ($E_T \approx E_i$) and $\sigma$s are the capture cross-sections. The quantity $\Gamma$, typically much larger than unity in a strong interfacial field, can be estimated by looking at the fractional change in emission coefficient $e_n = e_{n0}\exp{(E_C-E)/k_BT}$ in presence of a Boltzmann (Frenkel-Poole) jump over {a tilted} barrier and a WKB tunneling (Fowler-Nordheim) through the {tilted} barrier.  
\begin{eqnarray}
\Gamma &=& \displaystyle \int dE T(E)\frac{d(e_n/e_{n0})}{dE}\nonumber\\ &=&\frac{\Delta E_n}{k_BT}\int_0^1 exp\left[\frac{\Delta E_n}{k_B T}u-K_n u^{3/2}\right]du\\ \nonumber
K_n&=&\frac{4}{3}\frac{\sqrt{2m^*\Delta E_n^3}}{q\hbar F}
\end{eqnarray}
where $\Delta E_n = E_C-E_T$, $u = (E_C-E)/\Delta E_n$.
Assuming the trap density of states is localized at mid-gap, i.e., a delta function of weight $D_{it}(E_i)$, we then get the trap assisted current per unit width, under drain bias, to be \cite{SajjadCompactModel}
\begin{equation}
    I_{TAT}=\frac{q}{2} {v_{rcmb}n_i   \Gamma d}\left[1-e^{-\displaystyle qV_{DS}/{k_BT}}\right],
\end{equation}
where $d$ is the width of the trap active region along the transport direction, the recombination velocity $v_{rcmb}=\sigma v_{th} N_t$ and $\int D_{it}dE=\int N_t \delta(E-E_i) dE=N_t$. Here, $N_t$ represents the trap density per unit area at the midgap energy, while $D_{it}$ converts this into a density of states with a delta function profile at the trap energy.  

This TAT model can be incorporated in our analytical ballistic TFET model since it requires only the electric field at the junction.  The highest electric field $F$ can be evaluated from the potential model described in section \ref{sec:potential}. In this work we assume $\sigma = 5\times 10^{-17}$ $m^{2}$, $d=1$ $nm$ and $\Delta E_n=E_g/4$, because under the action of an electric field the trap level slips past the source conduction band edge and has a smaller barrier to the bandedge of the intrinsic region \cite{SajjadCompactModel}.  

\begin{figure}
\includegraphics[width=0.45\textwidth]{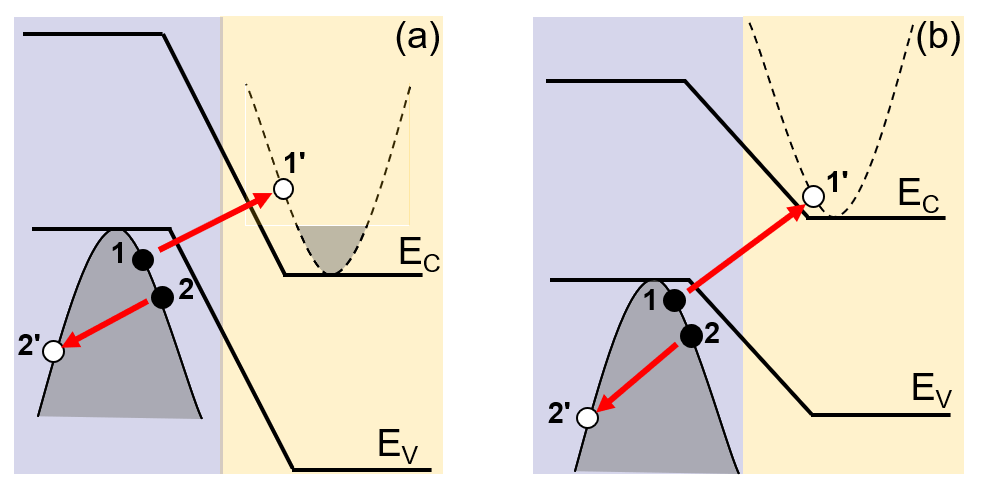}
\caption{Auger generation process of near the source/channel junction in (a) on and (b) off states in a TFET. }\label{fig:Auger_process}
\end{figure}
\subsection{Auger Current Model}
In our BTBT current model, we now add current due to Auger generation.
Auger involves charge} scattering through Coulumb interaction. In fact, {three}  particles exchange energy and momentum, as a consequence, one of the particles can transit from valence band to conduction band, as illustrated by Fig. \ref{fig:Auger_process}.  

To increase the ON current of TFETs, researchers have proposed near broken gap (almost type III) heterojunctions, where the conduction band of the {channel} lies just above the {valence} band of the source. The resulting narrow triangular barriers increase the ON current upon the onset of band-to-band tunneling. Besides the creation of interfacial traps due to unsatisfied bond valency at the hetero-interface, these triangular barriers tend to increase the wavefunction overlap due to the steepness of the potential (i.e., the built-in electric field),  which tends to shoot up the Auger generation (impact ionization) process depicted in Fig.~\ref{fig:Auger_process} (b) and increases the off current floor in the TFET. 
Through Auger generation process, extra holes in the source region and extra electrons in the channel region are generated. While the impact on the ON current (Fig.~\ref{fig:Auger_process} (a) is minor compared to the  BTBT term, any change in an otherwise low off current hurts the ON-OFF ratio and the subthreshold swing \cite{Teherani}.  
The rate of Auger generation   can be estimated by employing Fermi's Golden Rule
\begin{eqnarray}
    G & = & \frac{1}{A}\frac{2\pi}{\hbar}\sum_{1,1',2,2'}  P\left(1,1',2,2'\right) \nonumber\\
    & &|M|^2 \delta\left(E_1-E_{1'}+E_2-E_{2'}\right).
\end{eqnarray}
As depicted in Fig.~\ref{fig:Auger_process},  1 and 2 correspond to initial states of holes, while  1' and 2' correspond to final states in Auger generation process.
The P is the occupancy of the initial and final states, given by
\begin{eqnarray}
    P_{HCHH}(1,1',2,2') &=& \bar{f}_v(E_1)\bar{f}_v(E_2)f_c(E_{1^\prime})f_v(E_{2^\prime})
    \nonumber\\&-&f_v(E_1)f_v(E_2)\bar{f}_c(E_{1^\prime})\bar{f}_v(E_{2^\prime})\nonumber\\
    &\approx& \bar{f}_v(E_1)\bar{f}_v(E_2)f_c(E_{1^\prime}) - \bar{f}_v(E_{2^\prime})~~~~~~~~~~
\end{eqnarray}
where $\bar{f} = 1-f$ is the hole occupancy. The subscripts refer to the bands involved, heavy-hole and conduction band.
Within an envelope function approximation {for the matrix element $M$}, the Auger generation rate can be {written} as \cite{Teherani}
\begin{eqnarray}
    G &=& \frac{1}{A}\frac{4\pi}{\hbar}\sum_{1,1',2,2'} -\frac{p}{N_v} \exp\left(-\frac{E_{2'}-E_v}{kT}\right) \nonumber\\ 
    & &  \left(\frac{q^2}{2\epsilon A} \frac{  \delta_{k_{\perp 1}-k_{\perp 1'}+k_{\perp 2}-k_{\perp 2'}}}{|k_{\perp 1}-k_{ \perp 1'}|}\right)^2 \\
    & & (c_u K)^2 \left| \langle\psi_{1'} | \psi_1 \rangle \right|^2 \delta\left(E_1-E_{1'}+E_2-E_{2'}\right) \nonumber
\end{eqnarray}
In this equation,  $ |\psi_{1}\rangle $ and  $ |\psi_{1'}\rangle $ correspond to the envelope functions of the initial valence state $1$ and final conduction $1'$ as depicted in Fig.~\ref{fig:Auger_process}.
The $c_u K$  {involve the Bloch parts of the wavefunctions and are} evaluated using a 8 band $k\cdot p $ model\cite{Teherani}.  {For III-V semiconductors, $c_u$ is approximately $\sqrt{2\times10^{-17}}$ $cm$ and $K=\left|k_1-k_{1'}\right|$.}

The summation $\sum_{1,1',2,2'}$ has to be treated differently according to the device structure.
For bulk materials this summation sums over a twelve dimensional k-space since it involves four particles.
For quantum well TFETs studied  earlier \cite{Teherani}, 
the states $1$,$1'$, $2$ and $2'$ are quasi continuous in 2-dimensions, whereupon the $\sum_{1,1',2,2'}$ becomes
\begin{equation}
    \int d^2k_{\perp 1}d^2k_{\perp 2}d^2k_{\perp 1'}d^2k_{\perp 2'} \sum_{k_{x_1},k_{x_{1'}},k_{x_2},k_{x_{2'}}}
\end{equation}
Here  $k_{x_1}$,$k_{x_{1'}}$,$k_{x_2}$,$k_{x_{2'}}$ are discrete states due to quantum confinement  {in the wells in the transport direction}.
For planar TFETs considered in this work, the states $1$,$1'$, $2$ and $2'$ are quasi continuous in all 3-dimensions. The $\sum_{1,1',2,2'}$ becomes
\begin{equation}
    \int d^2k_{\perp 1}d^2k_{\perp 2}d^2k_{\perp 1'}d^2k_{\perp 2'} \int dk_{x_1}dk_{x_2}dk_{x_{1'}}dk_{x_{2'}} 
\end{equation} 
Since the device is considered as an infinite plane in y and z directions, $k_{\perp}$s are summed separately.
While the x direction is transport direction, the states $k_{x_1}$,$k_{x_{1'}}$,$k_{x_2}$,$k_{x_{2'}}$ are quasi-continuous states which have exponential tails near the source-channel junction. Therefore, the wave function overlaps $ \langle\psi_{1'} | \psi_1 \rangle $ depend on position x. 

To estimate the wave function overlap, we make use of the band model in \ref{sec:band_model}.
The wave vectors are calculated using equation (\ref{eq:k_x}) and the right decaying wave function of 1 is $\psi_{1} = \exp{\left( -\kappa_x (x-x_1) \right) }$ and left decaying wave function of 1'
 $\psi_{1'} = \exp{ \left( \kappa_x (x-x_1') \right)}$. 
Here we only consider the Auger generation at the junction, since the carriers (electrons) generated in the high-field region of a pn junction are swept out by the electric field to the channel.
Thereafter the current-density from Auger processes can be estimated as
\begin{equation}
    J_{aug} = q G
\end{equation}

{The Auger current can be obtained by enforcing momentum and energy conservation involving an electron and a hole (Fig.~\ref{fig:Auger_process})
\begin{eqnarray}
k_1 + k_2 &=& k^\prime_1 + k^\prime_2\nonumber\\
E_1 + E_2 &=& E^\prime_1 + E^\prime_2
\end{eqnarray}
We can eliminate $k_2$ and write $E^\prime_2$
in terms of $k_1$ and $k^\prime_1$ as independent variables ($k^\prime_2$ being set by $E^\prime_2$). We can then minimize $E^\prime_2$ with respect to $k_1$, $k^\prime_1$, which gives us $k_1=k_2$ and $J_{Aug} \propto \exp{\left(-E^\prime_{2,min}/kT\right)} = \exp{\left( -\frac{2\mu^{-1}+1}{\mu^{-1}+1}\frac{\Delta E}{kT}\right)}$ which defines the Auger generation limited sub-threshold}, where $\mu$ is the mass ratio $\mu = {m^*_c}/{m^*_v}$.
Here $\Delta E$ is the energy separating the lowest conduction band in the channel and the highest valence band in the source. The Auger generation limited sub-threshold ranges from $30$ {meV/decade when $m^*_c \ll m^*_v$ and $60$ meV/decade when $m^*_c \gg m^*_v$.}

{The Auger exponent is fairly easy to understand. If we make the effective mass $m^*_c$ very small, then we have a highly localized conduction band and $k^\prime_1 = 0$, in which case we get $k_1 = k_2 = k^\prime_2/2$ (co-moving electron and hole, or equivalently oppositely moving electrons in conduction band or holes in the valence band) and $E^\prime_{2,min} = 2\Delta E$. The double jump across the band-gap creates a very low subthreshold swing. On the other hand if $m^*_v \rightarrow 0$ then $k_1 = k_2 = 0$ and thus $k^\prime_2 = -k^\prime_1$ and $E^\prime_{2,min} = \Delta E$. Since the momentum is equally split, we are limited by the usual Boltzmann limit.}

\section{Results}
In our study, we simulated an InGaAs homojunction TFET and a GaSb/InAs heterojunction TFET. 
The homojunction and heterojunction TFETs {in our simulations} share the same device geometry, with the channel length $L_{ch}=100nm$, the channel thickness $t_{ch}= 5nm$, oxide thickness $t_{ox}=2nm$, and gate oxide dielectric constant of $\epsilon_r=11.9$. For the homojunction TFET, doping concentrations of $2\times10^{19}$  $cm^{-3}$, $10^{14}$ $cm^{-3}$ and $10^{18}$ $cm^{-3}$ are used for the source, channel and drain respectively. For the heterojunction TFET, doping concentrations of $5\times10^{19}$  $cm^{-3}$, $10^{14}$ $cm^{-3}$ and $5\times 10^{17}$ $cm^{-3}$ are used for the source, channel and drain respectively.

\begin{figure}[h] 
\includegraphics[width=0.45\textwidth]{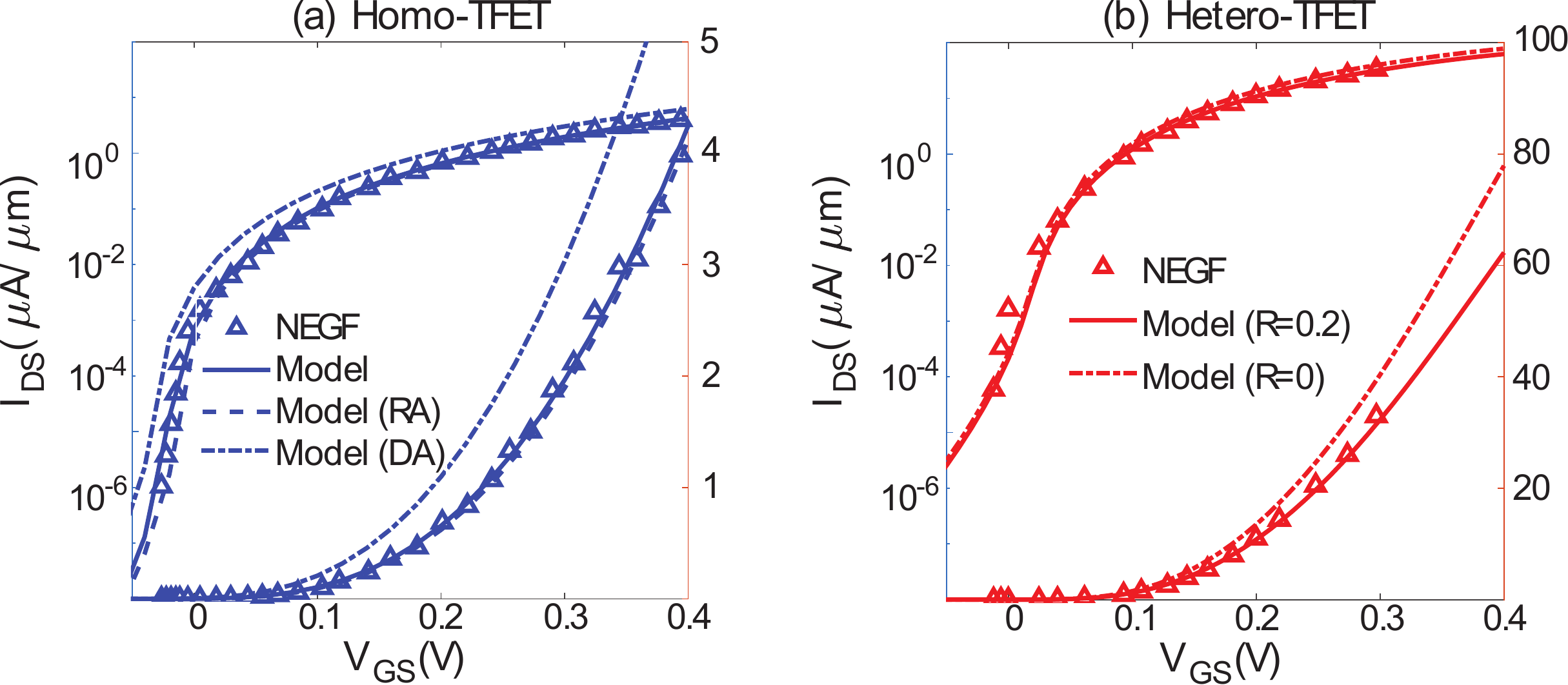}
\caption{$I_D-V_{GS}$ characteristics of (a) Homojunction and (b) Heterojunction TFETs by analytical model (this work) and NEGF\cite{Avci12,Long}. In (a), the results presented by solid lines are obtained by analytical models with material parameters extracted from the NEGF simulations for benchmark purpose. The dashed lines are by analytical models with material parameters extracted from tight binding calculations by this work, from Fig.~\ref{fig:model_complex_bands_random_digital_alloy}. In (b), the solid line uses $R=0$, corresponding to zero reflection at the InAs/GaSb interface; the dashed line uses $R=0.2$ to achieve better agreement with NEGF. {(Check line symbols for DA and RA in (a))}
 }\label{fig:TFET_benchmark}
\end{figure}

\begin{figure}[h] 
\centering
\includegraphics[width=0.4\textwidth]{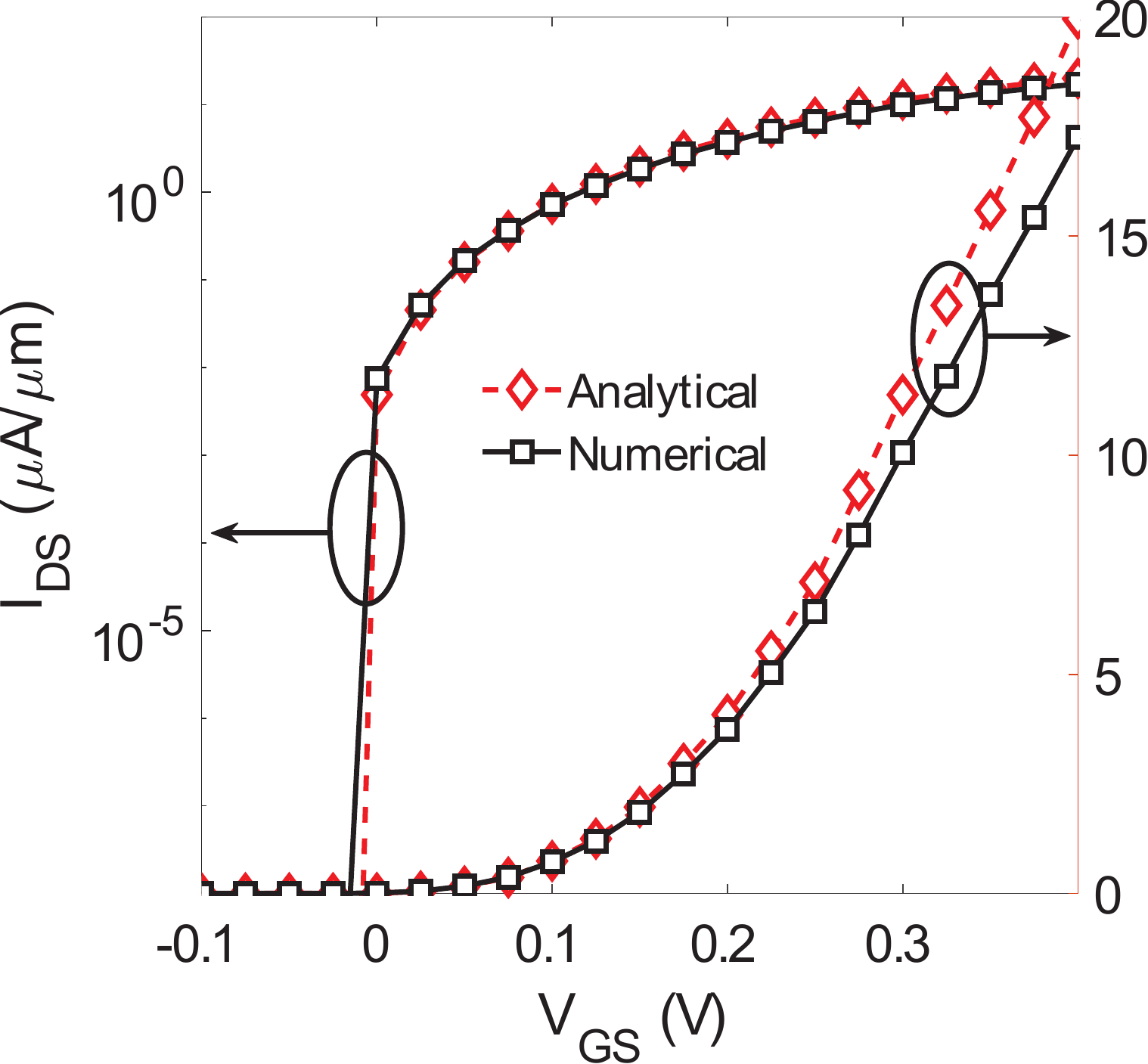} 
\caption{ Comparison of analytical vs. numerical calculation of the $I_D-V_{GS}$ characteristics of a homojunction TFET at $V_{DS}=0.3V$. The analytical result deviates from the numerical simulation by about $20\%$.  }\label{fig:anlytical_numerical_comparison}
\end{figure}
In order to benchmark the accuracy of our analytical model we compared the results of our model to previous tight binding based NEGF simulations {for ballistic TFETs}. The comparison is shown in Fig.~\ref{fig:TFET_benchmark}. Our analytical model of a 100nm homojunction TFET was compared to NEGF simulations of Avci {\it{et al.}} \cite{Avci12} while our 30nm heterojunction TFET data was compared to simulations carried out by Long {\it{et al}} \cite{Long}. For benchmarking purposes, we use  parameters such as $m^*$ and $E_g$ from past NEGF work, {with the understanding that at hetero-interfaces these parameters will change, and will need to be corrected later from our two-band model fitted to DFT}. The only free parameters we can adjust to calibrate our results for homo and heterojunctions are their gate work functions. The gate work function we used is 4.11 eV for the homojunction TFET and 4.642 eV for the heterojunction TFET.  Compared with the NEGF calculations, we see that our analytical model is in excellent agreement for the homo-junction TFET. We get a slightly higher current for the hetero-junction TFET if we assume no reflection at the InAs/GaSb interface ($R=0$). In reality, we expect some reflection at the material interface. To achieve a better agreement with NEGF for heterojunction TFET, we use a reflection parameter $R=0.2$. Our homo-junction ballistic TFET analytical model yields a steep subthreshold slope of 8.4$\textrm{mV/dec}$, which is slightly higher than the NEGF data which has a SS of 6.5 $\textrm{mV/dec}$ (some extraction error could exist); for heterojunction TFET, our analytical model leads to a SS of 14.9 $\textrm{mV/dec}$ which is in good agreement with the NEGF SS of 14.4 $\textrm{mV/dec}$. 

{Since the tunneling current depends exponentially on the material parameters, it is important to get these parameters accurately - which becomes questionable at interfaces for conventional tight-binding models that are typically fitted to bulk bandstructures and lack explicit atom-like localized non-orthogonal orbital basis sets. For comparison, we show}
in Fig.~\ref{fig:TFET_benchmark} (a) the $I_D-V_{GS}$ of the homojunction TFET using { material parameters extracted from our own tight binding calculations fitted to DFT (Fig.~\ref{fig:tightbinding_vs_DFT_2bandmodel})}. Based on  parameters of the random InGaAs alloy extracted from our tight binding calculation, we find that its $I_D-V_{GS}$ is higher than previous ballistic NEGF calculations with tight binding parameters fitted to experimental data. This is mainly due to the smaller band gap in the random alloy, as shown in table \ref{tab:material_parameters}. However, the digital alloy agrees with previous NEGF calculations. The difference in material parameters is summarized in table \ref{tab:material_parameters}:
\begin{table}
\centering
\begin{tabular}{cccccc}
\hline\hline
Material & InGaAs & InGaAs  & InGaAs & GaSb & InAs\\
 &(ref\onlinecite{Avci12} )& (random) & (digital)& (ref\cite{Long} ) & (ref\cite{Long} ) \\
\hline
$E_g (eV)$ & 0.740 & 0.704& 0.730& 1.20& 0.76\\
$m^* (m_0)$ & 0.041& 0.041& 0.043& 0.073& 0.052\\
\hline
\end{tabular}
\caption{Material parameters
}\label{tab:material_parameters}
\end{table}

\begin{figure}[h]
\includegraphics[width=0.45\textwidth]{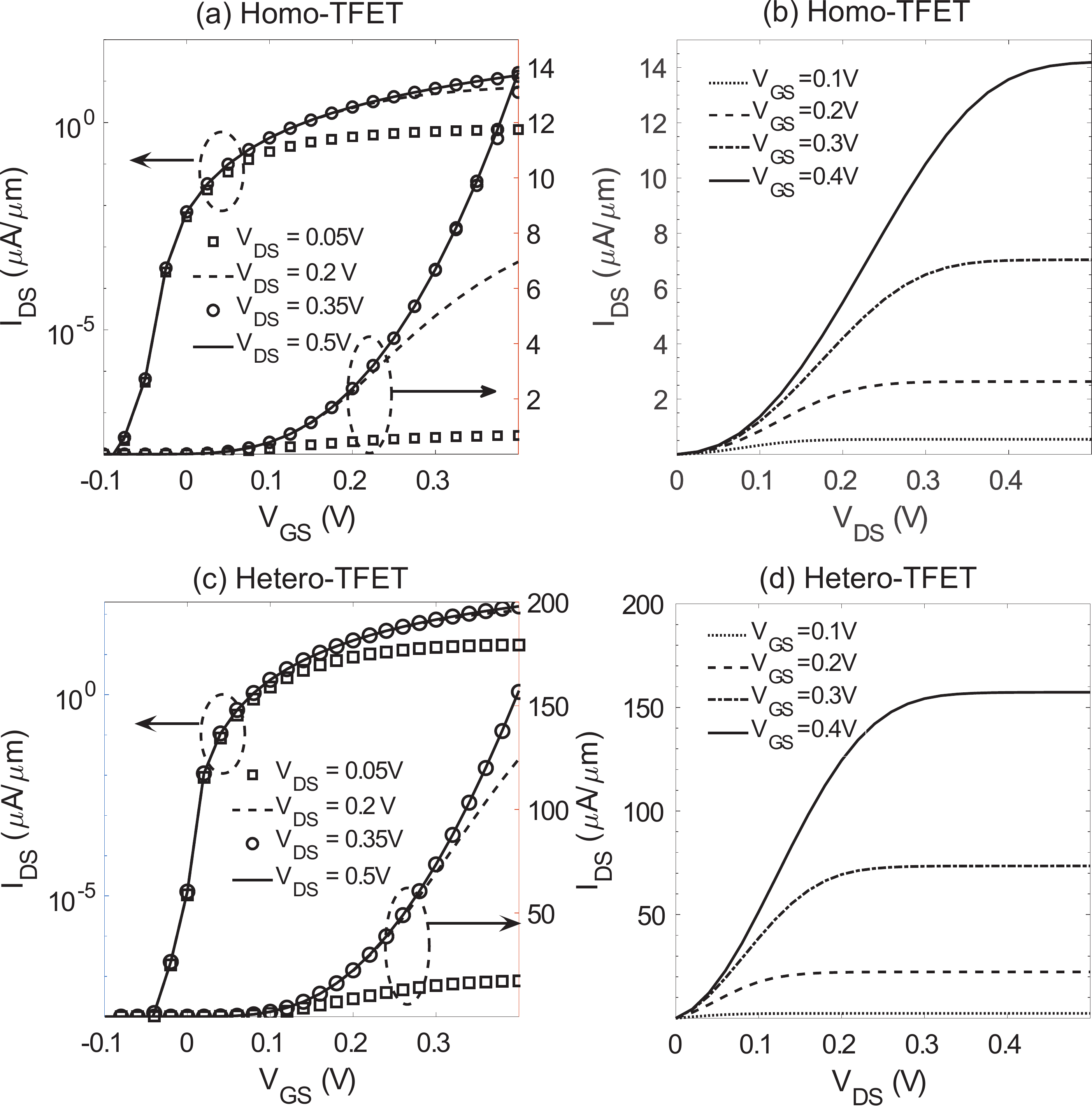}
\caption{(a) Transfer characteristics ($I_D-V_{GS}$) and (b) Output characteristics ($I_D-V_{DS}$) of Homojunction TFET; (c) Transfer characteristics ($I_D-V_{GS}$) and (d) Output characteristics ($I_D-V_{DS}$) of Heterojunction TFET
}\label{fig:Id_homo_hetero}
\end{figure}

Fig.~\ref{fig:Id_homo_hetero} shows the $I_D-V_{GS}$ and $I_D-V_{DS}$ of a {ballistic} homojunction TFET and heterojunction TFET. Here we used the 100nm channel for both homo and heterojunction TFETs. The heterojunction TFET shows a larger on-current for the same bias due to narrower tunneling barrier in the heterojunction TFET. For the homojunction TFET, we used the band gap and effective masses of random alloy InGaAs as shown in table \ref{tab:material_parameters}.
The $I_D-V_{GS}$ shows monotonous increasing behavior for both cases. The $I_D-V_{GS}$ shows a current saturation behavior for small $V_{DS}$, suggesting the integrated transmission in the TFET are saturated as well. The SS of  the homojunction TFET with 100nm channel length is 9.3 $\textrm{mV/dec}$, and the heterojunction TFET with 100nm channel length has a smaller SS of 6.8 $\textrm{mV/dec}$. {While these numbers are impressive, we will now see how non-idealities tend to affect these metrics.}

\begin{figure} 
\includegraphics[width=0.45\textwidth]{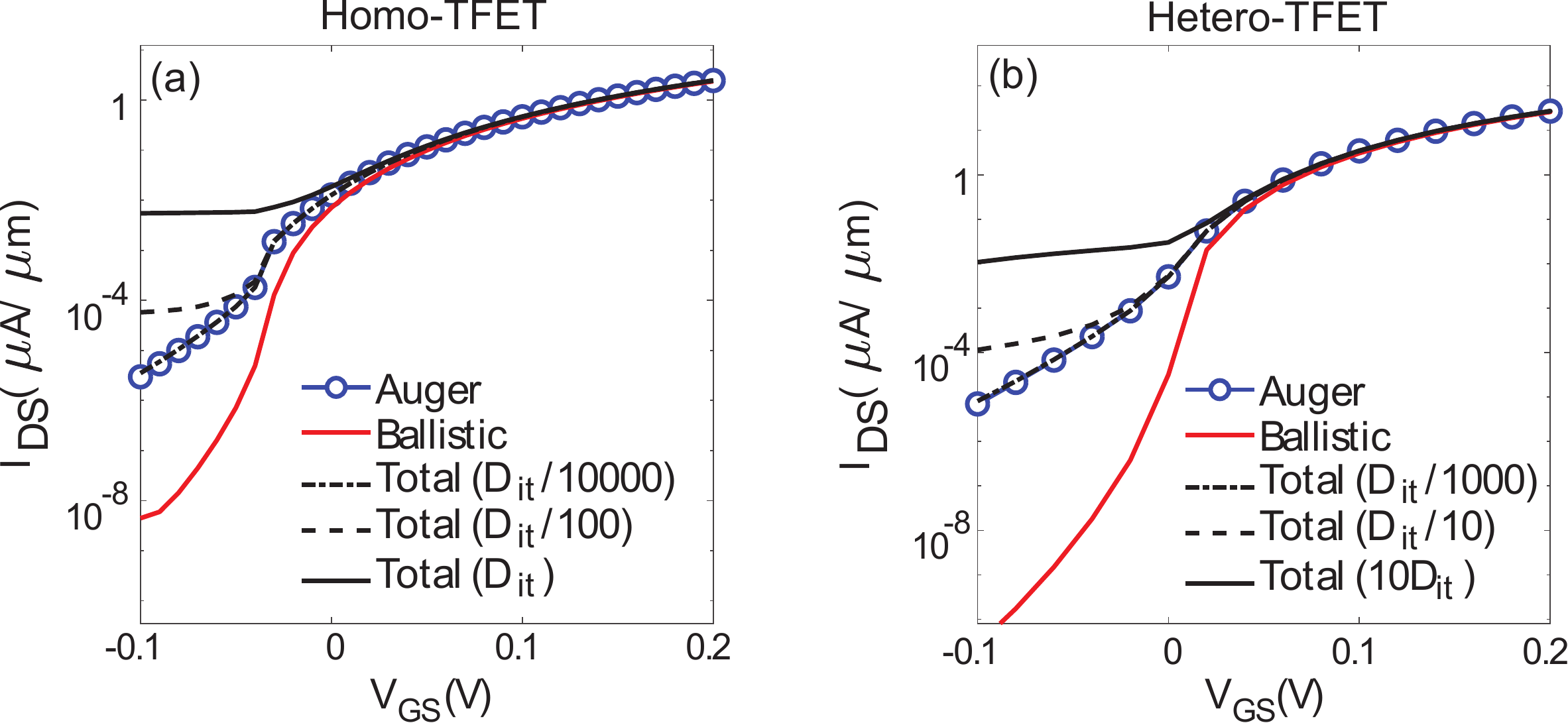}
\caption{ Comparison of non-ideal current with Ballistic, Auger and TAT current for Homojunction TFET. The Ballistic current is a steep switch, while the Auger effect increase the off-current and sub-threshold swing. The trap assisted tunneling current dominate the off-current when the density of trap is larger than $\frac{D_{it}}{ 10000} = 5\times 10^{12} m^{-2}eV^{-1}$.}\label{fig:non-ideal currents}
\end{figure}

Fig.~\ref{fig:non-ideal currents} shows the {\sout{ballistic}} $I_D-V_{GS}$ with various non-idealities such as trap assisted tunneling and Auger effect in homojunction and heterojunction TFETs. We see that the impact of the trap assisted tunneling process (TAT) is a significant increase of the off-current, in agreement with past simulations \cite{SajjadTraps}.

{Beyond trap assisted tunneling that is interface-specific, there is intrinsic leakage at high fields due to Auger generation that will also impact the off-current}.
In fact, Auger generation dominates the behavior of the off-current region when the traps are sufficiently low.
Our analysis shows that when the trap concentration at the junction of a  homo-TFET  is lower than a critical interface trap density of $5\times 10^{12} m^{-2}eV^{-1}$ {(trap density or trap density of states?? BIG difference.. industry is calibrated to the latter, not the former, so we need to translate to the latter number)} {$D_{it}$ is called the interface trap density in the literature (not only in Redwan's paper but in various books and paper I went through). However, it basically gives the trap density of states}, the Auger current begins to dominate the off-current region of the TFET. In contrast in hetero junction TFETs, the critical interface trap density is about one order of magnitude higher due to the intrinsically higher on-current. This Auger limited off-current defines a different limit of sub-threshold swing. Since the Auger generation current satisfies $J_{Aug} \propto \exp{\left( -\frac{2\mu^{-1}+1}{\mu^{-1}+1}\frac{\Delta E}{kT}\right)} $ , the Auger limited subthreshold swing is 
$\textrm{SS}_{Aug} \approx \frac{\mu^{-1}+1}{2\mu^{-1}+1} 60\textrm{mV/dec} $.

The Auger current depends on the wave function overlap of the electron and hole states across the junction.
In a TFET, this wave function overlap depends on the junction width and barrier height. 
In a heterojunction TFET, the junction width is smaller compared with homojunction TFET. 
However, barrier height also affect the wave function overlaps since the states decay faster in a higher barrier.
An ideal junction that minimize the Auger current is still an abrupt junction.

\subsection{ Impact on Circuit Energy}

To demonstrate the impact of Auger generation and trap assisted tunneling on digital circuit energy, we construct an analytical model based on device currents and generic circuit parameters. The energy consumption per clock cycle of a CMOS logic style digital circuit can be represented as a sum of the static and switching components
\begin{eqnarray}
    E_{switch}=Q_{switch}V_{DD} \label{Esw}\\
    E_{static}=\int_0^{t_{clk}}I_{static}V_{DD} \label{Estat}
\end{eqnarray}
where $Q_{switch}=C_{switch}V_{DD}=\alpha NC_g V_{DD}$ is the total amount of charge consumed from switching the transistors, $t_{clk}$ is the clock period given by
\begin{equation}
    t_{clk}=\frac{4C_gV_{DD}L_{dp}}{I_{on}}
\end{equation}
where $C_g$ is the gate capacitance of a single device and $L_{dp}$ is the critical path delay normalized to the delay of a single unit inverter. $I_{static}$ is the total static current due to inactive transistors which can be written as
\begin{eqnarray}
    I_{static}=I_{\textit{off}}N_{\textit{off}}
\end{eqnarray}
where $N_{\textit{off}}$ is the total number of inactive transistors. For a circuit with $N$ total transistors, $N_{\textit{off}}$ includes the transistors that are not switched during the clock cycle as well as the effective quantity of transistors that are switched and spend the rest of the clock cycle turned off
\begin{eqnarray}
    N_{\textit{off}}=\frac{N}{2}(1-\alpha)+\frac{\alpha N}{2}\left( 1-\frac{1}{L_{dp}} \right) \label{Noff}
\end{eqnarray}
where the activity factor $\alpha$ is the fraction of transistors that are switched each clock cycle, effectively representing how active the circuit is.
Combining \eqref{Esw} \textendash \eqref{Noff}, the total energy consumption per clock cycle can be expressed as
\begin{equation}
    E_{cycle}=\alpha NC_gV_{DD}^2 + 4C_gL_{dp}\frac{I_{\textit{off}}}{I_{on}} \frac{N}{2}\left(1-\frac{\alpha}{L_{dp}}\right)V_{DD}^2 \label{epc_eq}
\end{equation}
The on and off-currents $I_{on}$ and $I_{\textit{off}}$ correspond to device currents under the conditions
\begin{eqnarray}
    I_{on}=I\left(V_{GS}=V_{DS}=V_{DD}\right)\\
    I_{\textit{off}}=I\left(V_{GS}=0,V_{DS}=V_{DD}\right)
\end{eqnarray}
\begin{figure}[t!]  
\includegraphics[width=0.3\textwidth]{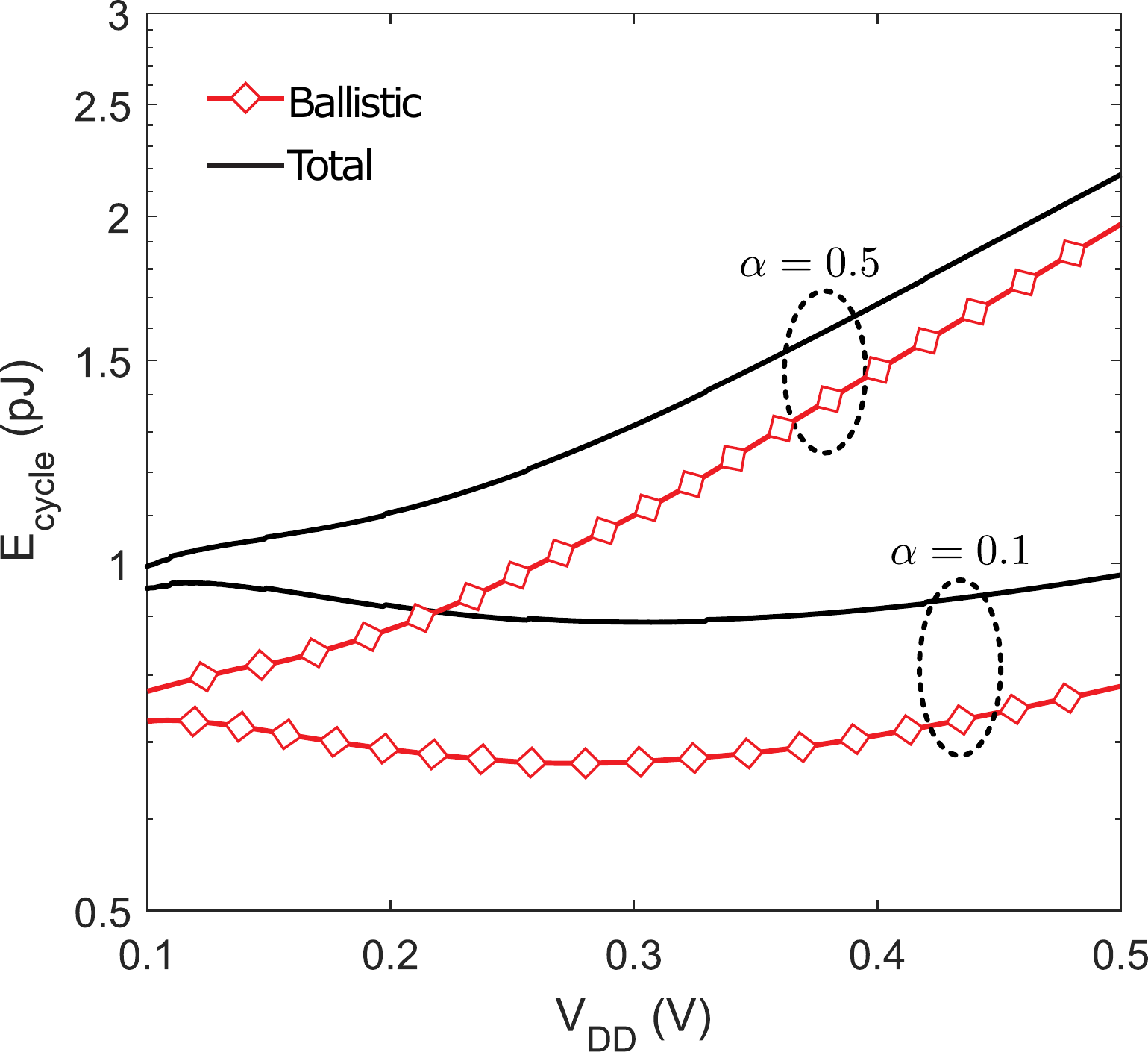}
\caption{Energy-Per-Cycle of a digital circuit with activity factor $\alpha$ of 0.1 and 0.5. Solid and marked curves are calculated using total device current (traps+Auger+ballistic) and ballistic-only currents respectively.}\label{fig:EPC}
\end{figure}
The energy per cycle of a digital circuit of $N=5000$ and $L_{dp}=80$ is calculated at $\alpha=0.5$ and $\alpha=0.1$, using \eqref{epc_eq} for both the ballistic current and the total current including Auger and trap-assisted tunneling for a heterojunction TFET. The results, shown in Fig.~\ref{fig:EPC}, demonstrate the increase in energy caused by the presence of Auger and trap-assisted tunneling currents. The energy increase is especially pronounced at low $V_{DD}$ and low $\alpha$, which is due to the fact that the energy in these regions is more heavily dependent on static current corresponding to $V_{GS}=0$. As shown in Fig.~\ref{fig:non-ideal currents}(b), this is where the Auger and trap-assisted tunneling currents have the biggest effect on total current compared to the ballistic current alone.
\section{Conclusion}
In this paper, we have presented an analytical model which captures the essential device physics of a TFET. The two-band $k.p$ model uses material parameters obtained from tight-binding complex band calculations fitted to $DFT$ at interfaces, and accurately represent the material bandstructure. The correct potential model and calibration of the model with NEGF simulations for both homo and heterojunction TFETs allow us to precisely  calculate the drain current at finite temperature, using a modified Simmons equation. The inclusion of trap-assisted tunneling and Auger generation processes into the model can explain the considerable observed discrepancy between ballistic vs experimental TFETs. Modeling of circuit energy further allows us to study the impact of these higher order effects on TFET circuit performance. Together, these tools can be used to understand the effects of higher order processes in TFETs and explore ways to mitigate their deleterious effects in order to improve performance of practical TFETs.

\begin{acknowledgments}
This work has been supported by the Semiconductor Research Corporation under Award 2694.02. 
\end{acknowledgments}

\providecommand{\noopsort}[1]{}\providecommand{\singleletter}[1]{#1}%

\end{document}